\documentclass[ floatfix]{revtex4}
\usepackage{amsmath}
\usepackage{color}
\usepackage{amssymb}
\usepackage{graphicx}
\usepackage[utf8]{inputenc}
\usepackage[english, russian]{babel}
\usepackage{textcomp}
\usepackage{color}
\usepackage{esint}

\begin{document}

\title{ Phase Diagram of UTe$_2$}

\author{V.P.Mineev}
\affiliation{Landau Institute for Theoretical Physics, 142432 Chernogolovka, Russia}

\begin{abstract}
The pressure-temperature phase diagram of superconducting UTe$_2$ with three lines of the second-order phase transitions cannot be explained in terms of successive transitions to superconducting states with a decrease in symmetry. The problem is solved using a two-band description of the superconducting state of UTe$_2$.
\end{abstract}
\date{\today}
\maketitle

\begin{figure}
\includegraphics
[height=.2\textheight]
{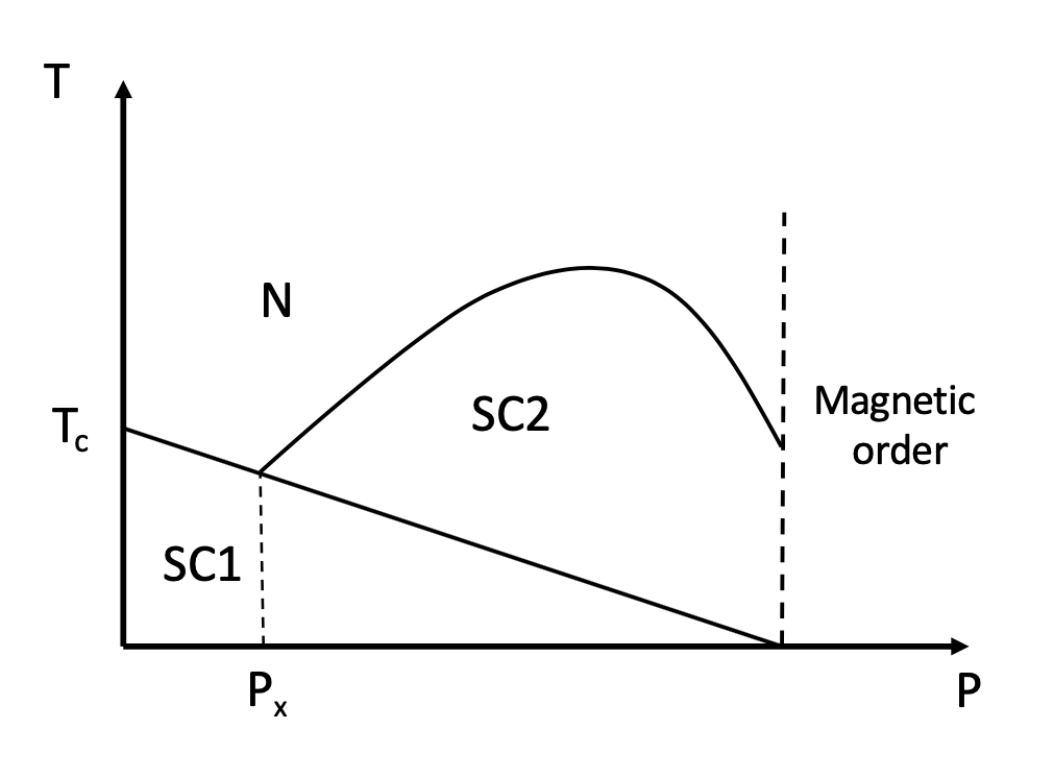}
 \caption{
Pressure-temperature phase diagram of UTe$_2$. }
\end{figure}

The recently discovered superconducting compound UTe$_2$  has many unusual properties (see review  \cite{Aoki2022} and references therein). It is a metal with an orthorhombic crystal structure and a superconducting critical temperature T$_c$=2.1 K. The transition temperature decreases with pressure, and then at  P$_x$=0.19 GPa the transition splits into two successive second-order transitions \cite{Braithwaite2019,Thomas2021} , as shown in Fig.1. A similar phase diagram is observed in a magnetic field along the b-crystallographic direction \cite{Rosuel2023,Vasina2024}. The extremely high upper critical field leaves no doubt that superconductivity in UTe2 is triplet-pairing superconductivity. The orthorhombic symmetry group D2h has four irreducible representations  $A,B_1,B_2,B_3$,  so in principle four superconducting states with order parameters
\begin{equation}
{\bf d}({\bf k})=\eta\left(\varphi_{x}(\hat{\bf k})\hat x+\varphi_{y}(\hat{\bf k}) \hat y+\varphi_{z}(\hat{\bf k}) \hat z\right)
\end{equation}
are possible. The particular form of the functions $\varphi_{x},\varphi_{y},\varphi_{z}$ corresponds to the symmetry of a specific representation. As usual, the axes x, y, z are directed along the crystallographic directions $a, b, c$. The phase transition lines of two of them with different pressure dependences can intersect, so that the superconducting phase arising below both transition lines must have a symmetry group that is a subgroup of the symmetry groups of both initial superconducting phases  \cite{LL}. A phase diagram of this type inevitably contains four lines of second-order phase transitions. However, the phase diagram shown in Fig. 1, containing only three lines, is completely different. Therefore, an explanation based on successive phase transitions to superconducting states with a decrease in symmetry does not correspond to reality. 

An explanation of a phase diagram with three lines of second-order transitions can be found in the model of two band superconducting states belonging to the same representation. Let us discuss the situation with a magnetic field along the b-crystallographic direction ${\bf H}=H\hat y$.  The critical temperature in the absence of orbital effects is determined by a system of linear equations for the order parameters $\eta_1,\eta_2$

\begin{eqnarray}
\eta_i=\frac{1}{2}\int\frac{d^3{\bf k}}{(2\pi)3}\sum_{\omega_n}\sum_jV_{ij}\eta_j\left\{(\phi_{xj}^2(\hat{\bf k})+\phi_{zj}^2(\hat{\bf k}))
\left [G^+_j({\bf k},\omega_n)G^+_j(-{\bf k},-\omega_n)+G^-_j({\bf k},\omega_n)G^-_j(-{\bf k},-\omega_n)\right]+\right.\nonumber\\
\left.+\phi_{yj}^2(\hat{\bf k})
\left[G^+_j({\bf k},\omega_n)G^-_j(-{\bf k},-\omega_n)+G^-_j({\bf k},\omega_n)G^+_j(-{\bf k},-\omega_n)\right]\right\},
\label{2}
\end{eqnarray}
where, $ i=1,2$, $ j=1,2$ are the band indexes. $V_{ij}$ is the matrix including the  amplitudes of intraband pairing $V_{11}$, $V_{22}$ and interband pair scattering  $V_{12}=V_{21}$. The normal state electron Green functions are
\begin{equation}
G^\pm_i({\bf k\omega_n}=\frac{1}{i\omega_n-\xi_i\pm\mu_BH}
\label{3}
\end{equation}
and
\begin{equation}
\xi_i=\varepsilon_i({\bf k})-\mu.
\end{equation}
The second line in Eq.(\ref{2}) corresponds to paramagnetic suppression of $y$ component of the order parameter by magnetic field. Equations 
(\ref{2}) as well the Equation (\ref{11}) below are  written in neglect tiny effect of mixing of the order parameters  from different representations  resulting in formation of nonunitary superconducting states.

In the field absence, multiplying Eq.(\ref{2})  on $V_{ij}^{-1}$ and performing necessary summations and integration we come to the system of equations
\begin{eqnarray}
\left(\frac{V_{22}}{D}-\bar N_1\ln\frac{\varepsilon_0}{T_c}\right)\eta_1-\frac{V_{12}}{D}\eta_2=0,
\label{5}\\
-\frac{V_{21}}{D}\eta_1+\left(\frac{V_{11}}{D}-\bar N_2\ln\frac{\varepsilon_0}{T_c}\right)\eta_2=0,
\label{6}
\end{eqnarray}
where $D=V_{11}V_{22}-V_{12}^2$, 
$\bar N_i=\left \langle \left( \phi_{xi}^2(\hat{\bf k})+\phi_{yi}^2(\hat{\bf k})+\phi_{zi}^2(\hat{\bf k})\right)N_{0i}(\hat{\bf k})\right \rangle$ is the density of states averaged over corresponding Fermi surface, 
$\varepsilon_0$ is a cut-off energy for pairing interaction.  Equating the determinant of the system  to zero, we obtain the transition temperature
\begin{equation}
T_c=\varepsilon_0\exp\left(-\frac{1}{\lambda}\right),
\label{T_c}
\end{equation}
where
\begin{equation}
\lambda=\lambda_{\pm}=(\lambda_{11}+\lambda_{22})/2\pm\sqrt{(\lambda_{11}-\lambda_{22})^2/4+\lambda_{12}\lambda_{21}},
\end{equation}
and $\lambda_{ij}=V_{ij}\bar N_j$. 

So long $V_{12}\ne 0$, the superconducting state in two bands has common critical temperature 
with 
\begin{equation}
\lambda=\lambda_+.
\end{equation}
When $V_{12}=0$ the superconducting states in two bands are formed independently with critical temperatures determined by Eq.
(\ref{T_c}) with
\begin{equation}
\lambda=\lambda_{11}~~~~~~~~\text{and}~~~~~~~~\lambda=\lambda_{22},
\end{equation}
correspondingly.
All three parameters  $\lambda_{11}$,  $\lambda_{22}$, $\lambda_{12}$ depend from pressure. 
If at some pressure $P_x$ the interband pair scattering vanishes then the superconducting state in each band appears at its own critical temperature $T_{c\pm}$ in correspondence to signs $\pm$ in front of square root. To be consistent with behaviour shown on Fig.1 at pressures near  $P_x$ the parameters $\lambda_{11}$  and  $\lambda_{22}$ should be 
almost equal each other. See Fig.2.

\begin{figure}
\includegraphics
[height=.2\textheight]
{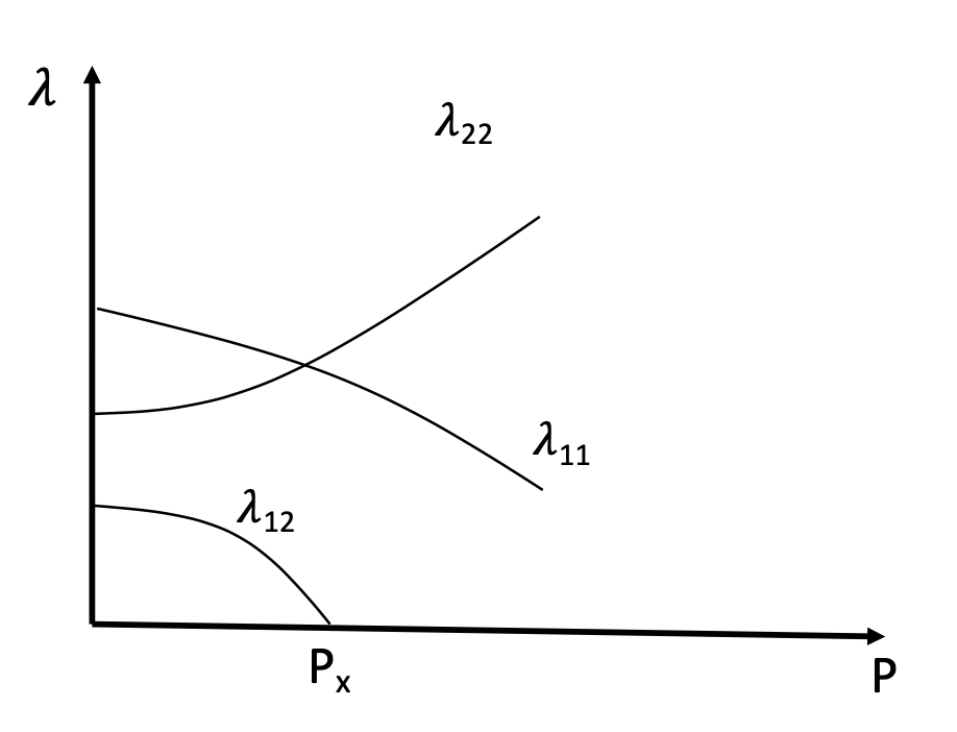}
 \caption{
Pressure dependence of parameters $\lambda_{11}=\lambda_{11}(P)$, $\lambda_{22}=\lambda_{22}(P)$, and $\lambda_{12}=\lambda_{12}(P)$. See text. }
\end{figure}

Obviously, this is a rather naive description of two-band superconductivity in the weak-coupling theory. In the strong-coupling theory, instead of the parameters $\lambda_{ij}$, we deal with the spectral density of boson excitations responsible for pairing. The amplitude of interband pair scattering will be expressed through the spectral weight of excitations with interband wave vectors. In strong coupling, the transition from the single-transition regime to the double-transition regime is not so sharp. The physical reasons for the proposed dependences are currently unknown. Their search is a task for future research.

In the presence of a magnetic field, taking into account both the Zeeman effect and the orbital effect, one can derive the Ginzburg-Landau equations, generalizing equations (\ref{5}), (\ref{6}). The solution of these equations can only be found numerically or using a variational procedure, described, for example, in the paper \cite{Dao2004}. We will not reproduce here this procedure, which is necessary at pressures P<P$_x$.

The problem becomes simpler for P>P$_x$, where $V_{12}=0$.
In addition, according to recent NMR data \cite{Kinjo2023}, the spin susceptibility along the $b$-axis remains unchanged in SC2, which differs from that observed in SC1 at ambient pressure and below the SC2-SC1 transition. This observation means that the function $\phi_{y2}(\hat{\bf k})=0$ and the Zeeman effect can be ignored in the GL equation
\begin{equation}
\left(\frac{1}{V_{22}\bar N_2}-\ln\frac{\varepsilon_0}{T}\right)\eta_2-K_x\left(\frac{2eHz}{\hbar c}   \right )^2\eta_2+K_y\frac{\partial^2\eta_2}{\partial y^2}+K_z\frac{\partial^2\eta_2}{\partial z^2}=0,
\label{11}
\end{equation}
where $\bar N_2=\left\langle \left( \phi_{x2}^2(\hat{\bf k})+\phi_{z2}^2(\hat{\bf k})\right)N_{02}(\hat{\bf k})\right\rangle$,  $K_a=\frac{7\zeta(3)\hbar^2}{32\pi^2T_{c2}^2\bar N_2}\left\langle\left( (\phi_{x2}^2(\hat{\bf k})+\phi_{z2}^2(\hat{\bf k})\right)v^2_{F2a}({\bf k})N_{02}(\hat{\bf k})\right\rangle$,  $a=x,y,z$.
The upper critical field is
\begin{equation}
H_{c2}= \frac{\hbar c}{2e}\frac{1}{\sqrt{K_xK_z}} \frac{T_{c2}-T}{T_{c2}}.
\end{equation}

In summary, the developed two-band approach to the superconducting state of UTe$_2$ allows to explain the strange pressure-temperature phase diagram
with three lines of second-order phase transitions. The physical origin of the supposed dependence of the parameters $\lambda_{ij}$ on pressure is currently unknown. Its explanation is a task for future research.

As already mentioned, the phase diagram with three lines of the second-order phase transition in UTe$_2$ also occurs at ambient pressure in a magnetic field along the $b$-crystallographic direction \cite{Rosuel2023,Vasina2024}. An explanation of such behaviour can also be obtained within the framework of the two-band description of the superconducting state.

I am indebted to M.E.Zhitomirsky and J.Flouquet for the interest to work and to J.-P.Brison for useful discussion.

\end{document}